\documentclass[aps,prd,onecolumn,amssymb]{revtex4}
\usepackage{graphicx}
\input epsf
\usepackage{graphicx,bm,color}
\usepackage[english]{babel}
\usepackage{amsmath}
\usepackage{amsfonts}
\usepackage{color}

\begin{document}

\font\bss=cmr12 scaled\magstep 0

\title{Some models of holographic dark energy on the Randall-Sundrum brane and observational data}
\author{Artyom V. Astashenok, Alexander S. Tepliakov\\
\small Immanuel Kant Baltic Federal University\\
\small Department of Physics, Technology and IT\\
\small 236041 Kaliningrad, Russia, Nevskogo str.14}

\begin{abstract}
The some models of holographic dark energy for Randall-Sandrum
brane are considered.  For first class of dark energy models we
take energy density in form $\sim L^{2\gamma-4}$ where $L$ is size
of events horizon in Universe and $\gamma$ is parameter (Tsallis
holographic energy). Analysis of observational data allows to
define upper limit on value of $\delta=\rho_{0}/2\lambda$
($\rho_{0}$ is current energy density in the Universe and
$\lambda$ is brane tension). Then we investigate models for which
dark energy density has form
$\rho_{de}=C^{2}L^{-2}-C_{1}^{2}H^{2}$ where $H$ is Hubble
parameter.
\end{abstract}
\medskip

 \maketitle

\section{Introduction}

From moment of discovery of cosmological acceleration in 1998
~\cite{1,2} explanation of this fact became one of the most
puzzles for theoretical physics and cosmology. The main way for
resolution of this ambiguous task is postulate the existence of so
called dark energy with very unusual properties. The parameter of
state $w=p/\rho$ for dark energy is negative. According to most
simple and successful model dark energy is nothing else than
Einstein cosmological constant or vacuum energy and its density
consist of  $~$70\% of total energy density in the Universe
\cite{LCDM-1}-\cite{LCDM-7}.

It is assumed that ultimate resolution of dark energy problem will
be achieved in frames of quantum gravity. An interesting approach
to this task is related to holographic principle. According to
holographic principle all physical quantities inside Universe
including dark energy density can be described by some values on
spacetime boundary ~\cite{3,4,5}. There are only two parameters
through which one can calculate dark energy density, Planck mass
$M_p$ and some characteristic lenghtscale $L$ namely
\begin{equation}
    \label{eq:1}
    \rho_{de}=3C^{2}M_{p}L^{-2}.
\end{equation}
For $L$ one can take for example size of events horizon, particles
horizon or inverse value of Hubble parameter $H$:
$$
L_{e}=a(t)\left ( \int_{t}^{\infty }\frac{d{t}'}{a(t)} \right ),
$$
\begin{equation}
\label{eq:3} L_{p}=a(t)\left ( \int_{0}^{t}\frac{d{t}'}{a(t)}
\right),
\end{equation}
$$
L_{h}=\frac{1}{H}.
$$
The various aspects of holographic dark energy are investigated in
many papers (see \cite{Nelson}-\cite{Wang} and reference therein).
As shown model with Hubble horizon are not suitable for
description cosmological evolution of our universe. The main line
of investigations is using size of events horizon as infrared
cut-off. It is interesting to note that holographic principle can
be applied to early universe too \cite{Nojiri-2} with obtaining of
inflation scenario. The cosmological bounce from holographic
principle was considered in \cite{Nojiri-3}.

In 1988 K. Tsallis proposed the generalized expression for black
hole entropy $S_{BH}$\cite{Tsallis}:
\begin{equation}
    \label{eq:6}
    S_{BH}= \mu A^{\gamma}.
\end{equation}
Here $\mu$ is unknown constant, $\gamma$ is non-additivity
parameter and $A$ is black hole horizon area. It is obviously that
well-known Bekenstein entropy
$$
S_{BS}=\frac{A}{4}
$$
follows from equation (\ref{eq:6}) if one put $\gamma$ = $1$ and
$\mu$ = $\frac{1}{4}$. If one assume that such approach is
suitable for dark energy then its density can be written as
~\cite{9}:
\begin{equation}
    \label{eq:7}
    \rho_{de}=C^{2}L^{2\gamma-4}.
\end{equation}
The model of Tsallis holographic dark energy for Friedmann
universe was proposed recently \cite{Bamba} for describing of
late-time acceleration. Authors of \cite{Nojiri-4} considered the
generalization of Tsallis holographic dark energy model.

In the simplest holographic dark energy model ($\gamma=1$) the
value of $C$ is around $0.75$ and universe ends its life in big
rip singularity. There are several approaches for resolving
problem of singularity have been proposed. One of them is brane
world scenario \cite{Brane}. According to Randall-Sundrum model
brane is our 4-dimensional Universe with infinitely thin wall
located in 5-dimensional spacetime ~\cite{7,8}. All fields of
Standard Model ``lives'' only on brane except gravity which can be
appear in additional dimensions. The cosmological equation on the
brane change their form in comparison with the standard Friedmann
cosmology namely:
\begin{equation}
\label{eq:5} H^{2}=\rho\left (1+\frac{\rho }{2\lambda } \right),
\quad H^{2}\equiv\frac{\dot{a}}{a},
\end{equation}
where $\rho$ is energy density, $a$ is scale factor and $\lambda$
is brane tension. Hereinafter we use natural system of units
($8\pi G/3=c=1$). The dependence of energy density $\rho$ from
scale factor can be obtained from the equation:
\begin{equation}
\dot{\rho}+3H(\rho+p)=0.
\end{equation}

In brane cosmology the size of events horizon tends to
$L_{e}\rightarrow L_{0}\neq 0$ for $t\rightarrow\infty$. Therefore
the density of holographic energy tends to constant value and we
have effective $\Lambda$CDM model.

In paper we considered the holographic dark energy model in
Tsallis form on Randall-Sandrum brane for varios values of
$\gamma$. We take $1<\gamma<2$ and compare results with case of
``ordinary'' holographic dark energy ($\gamma=1$). We investigated
the future evolution of the universe. Using observational data
allows to define limit on possible relation of current energy
density to brane tension $\rho_{0}/2\lambda$. We considered the
dependence between apparent magnitude and redshift for distant
supernovae Ia, Hubble parameter for some redshifts and baryon
acoustic oscillations. In wide range of parameters these data are
described well but only separately. For relatively small brane
tensions there are no common parameters at which all data are
satisfied with good accuracy. Finally we studied model in which
besides classical holographic contribution to dark energy $\sim
L_{e}^{-2}$ term $~H^{2}$ appears.

\section{Tsallis holographic dark energy model on Randall-Sandrum brane}

For our analysis we assume that spatially flat Universe is filled
by cold matter and dark energy only (contribution of radiation and
other components is negligible), i.e.
$$
\rho=\rho_{de}+\rho_{m}.
$$
For dark energy density we choose following representation:
\begin{equation}\label{Tsal}
\rho_{de}=\frac{C^{2}}{L_{e}^{4-2\gamma}}.
\end{equation}
Let's investigate this cosmological model for some values of ratio
between current energy density and tension of the brane
$\delta=\rho_{0}/2\lambda$. For this purpose it is convenient to
rewrite energy density $\rho_0$ via Hubble parameter:
$$
\rho_0=H_{0}^{2}(1+\delta)^{-1}.
$$
One can introduce dimensionless units for Hubble parameter,
density and brane tension by following
$$
H\rightarrow H_{0} \tilde{H}, \quad \rho\rightarrow
\rho_{0}\tilde{\rho},\quad \lambda\rightarrow\tilde{\lambda}
H_{0}^{2}.
$$
Then rewrite first Friedmann equation on the brane in the
dimensionless form:
\begin{equation}\label{FR1}
H^{2}=(1+\delta)^{-1}(\rho_{de}+\rho_{m})\left(1+\delta(\rho_{de}+\rho_m)\right).
\end{equation}
Here tildes are omitted.  One consider the
$\Omega_{de}=\rho_{de}/\rho_{0}$ and constant $C$ as varying
parameters. Therefore fraction of matter energy density is equal
${1-\Omega_{de}}$ and dimensionless matter density can be written
as
$$
\rho_{m}=\frac{1-\Omega_{de}}{a^{3}}.
$$
For our moment of time one can put $a(0)=1$ without loss of
generality. Differentiation on time of $L_{e}/a$ gives following
equation:
\begin{equation}
\frac{d(L_{e}/a)}{dt}=-\frac{1}{a(t)}. \label{DL}
\end{equation}
Equations (\ref{FR1}), (\ref{DL}) consist of first-order system of
differential equations for scale factor $a(t)$ and function
$L_{e}(t)/a(t)$. For condition on $L_{e}(0)$ one need define such
value that dimensionless energy density for dark energy is
$\Omega_{de}$ therefore
$$
L_{e}(0)=\left(\frac{C^2}{\Omega_{de}}\right)^{\frac{1}{4-2\gamma}}
$$
Acceptable models should describe astrophysical data with good
accuracy. We use standard statistical approach for estimation of
likelihood of cosmological model with some parameters namely
$\chi^2-$criteria. The following data are included in our
consideration.

1) \textbf{The dependence magnitude - redshift for supernovae Ia}.
The theoretical value of apparent magnitude $\mu_{t}$ for
supernova Ia with redshift $z$ can be calculated using formula:
\begin{equation}
\label{eq:12} \mbox{\bf $\mu$}_{th} = 5
\log_{10}\left[\frac{d_L(z)}{Mpc}\right] + 25
\end{equation}
The luminocity distance ${d}_L$ for spatially flat Universe is
\begin{equation}
\label{eq:13} {d}_L = \frac{c}{H_{0}}\int_0^z\frac{dz'}{E(z')}
\end{equation}
where $E(z)=H(z)/H_{0}$ is dimensionless Hubble parameter. For
$\chi^2_{SN}$ we have simple expression:
\begin{equation}
\label{eq:15} \chi _{SN}^{2}=\sum_{i}\frac{(\mu _{obs}(z_{i})-\mu
_{t}(z_{i}))^{2}}{\sigma _{i}^{2}}.
    \end{equation}
Here $\sigma_i$ is value of error for given measurement. We use
data on supernovae Ia given in ~\cite{9}.

2) \textbf{Baryon acoustic oscillations.} For these data it is
important to calculate so called acoustic parameter $A(z)$.
Theoretical value of acoustic parameter is given by
\begin{equation}
\label{eq:16} A_{t}(z)=\frac{D_{v}(z)H_{0}\sqrt{\Omega_{mo}}}{z},
\end{equation}
where $D_{v}(z)$ is distance parameter defined by relation:
\begin{equation}
\label{eq:17} D_{v}(z)=\left \{
(1+z)^{2}d_{A}^{2}(z)\frac{cz}{H(z)} \right \}^{1/3}.
\end{equation}
Here $d_{A}(z)$ is angular diameter distance:
\begin{equation}
\label{eq:18} d_{A}(z)=\frac{y(z)}{H_{0}(1+z)}, \quad
y(z)=\int_{0}^{z}\frac{dz}{E(z)}.
    \end{equation}
The parameter $A_{t}(z)$ can be evaluated via dimensionless
quantities:
\begin{equation}
\label{eq:19} A_{t}(z)=\sqrt{\Omega_{mo}}\left
(\frac{y^{2}(z)}{z^{2}E(z)}\right )
\end{equation}
For $\chi _{A}^{2}$ we calculate by the following way:
    \begin{equation}
    \label{eq:20}
    \chi _{A}^{2}=\Delta \mathbf{A}^{T}(C_{A})^{-1}\Delta \mathbf{A}
    \end{equation}
where $\Delta \mathbf{A}$ is vector with components $\Delta
A_{i}=A_{t}(z_{i})-A_{obs}(z_{i})$ and $(C_{A})^{-1}$ is inverse
matrix to covariance matrix $3\times 3$, which elements are given
in table \ref{tab:1}.
\begin{table}
\caption{\label{tab:1}Observed values of acoustic parameter for
various redshifts from ~\cite{11}}
\begin{center}
\begin{tabular}{|c|c|c|}
\hline $z$ & $A(z)$ & $\sigma _{A}$ \\
\hline 0.44 & 0.474 & 0.034 \\
\hline 0.60 & 0.442 & 0.020 \\
\hline 0.73 & 0.424 & 0.021 \\
\hline
\end{tabular}
\end{center}
\end{table}

3) \textbf{The dependence of Hubble parameter from redshift.} The
past evolution of Hubble parameter from time is studied
sufficiently well now. The Hubble parameter can be defined from
relation:
\begin{equation}
\label{eq:21} dt=-\frac{1}{H}\frac{dz}{1+z}.
\end{equation}
Therefore definition of $dz/dt$ allows to measure $H(z)$ directly.
These measurements are possible due to data about ages of galaxies
determined from star population models. The theoretical dependence
of Hubble parameter from redshift can be defined as
\begin{equation}
H(z)=H_{0}E(z), \quad
E(z)=\left(\rho(z)/\rho_{0}\right)^{1/2}.\label{eq:22}
\end{equation}
For Randall-Sandrum brane one need to slightly modify this
relation:
\begin{equation}
\label{eq:23} H(z)=H_{0}E(z)(1+\delta E(z))^{1/2}(1+\delta)^{-1/2}
\end{equation}
The value of $\chi_{H}^{2}$ is equal to
\begin{equation}
\label{eq:24}
\chi_{H}^{2}=\sum_{i}\frac{(H_{obs}(z_{i})-H_{t}(z_{i}))^{2}}{\sigma_{i}^{2}}
\end{equation}
The data about $H(z)$ are taken from ~\cite{12} and presented in
table \ref{tab:2}.
\begin{table}
\caption{\label{tab:2}The dependence of Hubble parameter $H$
(km/s/Mpc) from redshift $z$.}
\begin{center}
\begin{tabular}{|c|c|c|c|c|c|c|c|c|c|c|c|}
\hline $z$ & $H(z)$ & $\sigma_{H}$ & $z$ & $H(z)$ & $\sigma_{H}$
& $z$ & $H(z)$ & $\sigma_{H}$ & $z$ & $H(z)$ & $\sigma_{H}$\\

\hline 0.070 & 69 & 19.6 & 0.270 & 77 & 14 & 0.593 & 104 & 13 & 0.900 & 117 & 23 \\

\hline 0.090 & 69 & 12& 0.280 & 88.8 & 36.6 & 0.600 & 87.9 & 6.1 & 1.037 & 154 & 20 \\

\hline 0.120 & 68.6 & 26.2 & 0.350 & 76.3 & 5.6 & 0.680 & 92 & 8 & 1.300 & 168 & 17 \\

\hline 0.170 & 83 & 8 & 0.352 & 83 & 14 & 0.730 & 97.3 & 7 & 1.430 & 177 & 18 \\

\hline 0.179 & 75 & 4 & 0.400 & 95 & 17 & 0.781 & 105 & 12 & 1.530 & 140 & 14 \\

\hline 0.199 & 75 & 5 & 0.440 & 82.6 & 7.8 & 0.875 & 125 & 17 & 1.750 & 202 & 40 \\

\hline 0.200 & 72.9 & 29.6 & 0.480 & 97 & 62 & 0.880 & 90 & 40 & 2.300 & 224 & 8 \\

\hline
\end{tabular}
\end{center}
\end{table}

\begin{figure}[htbp]
    \centering
    \includegraphics [width=0.35\textwidth]{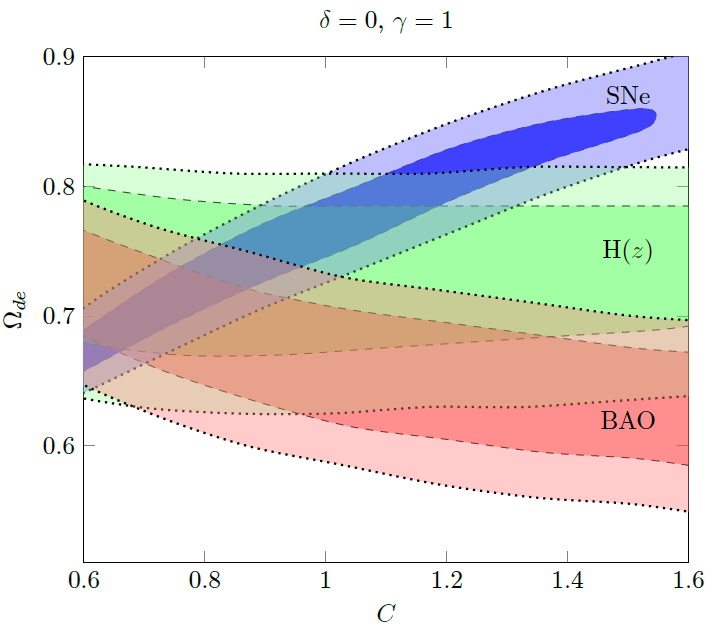}\includegraphics [width=0.35\textwidth]{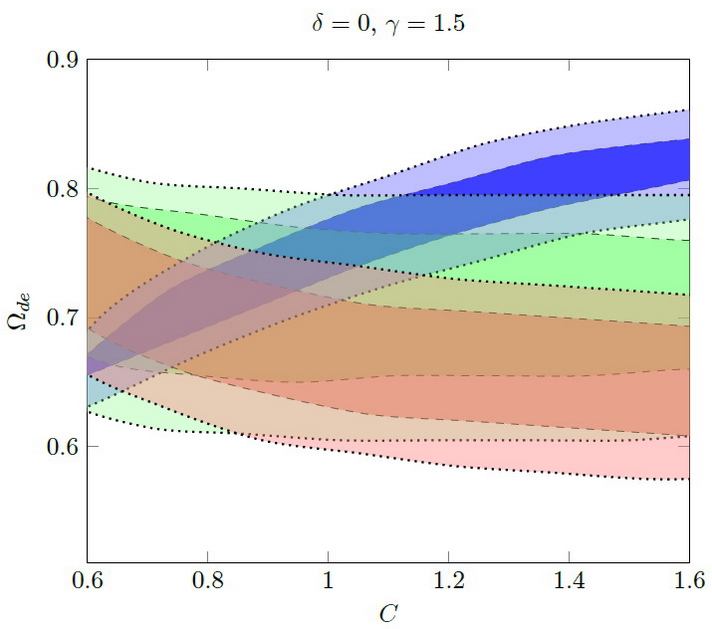}\\
    \includegraphics[width=0.35\textwidth]{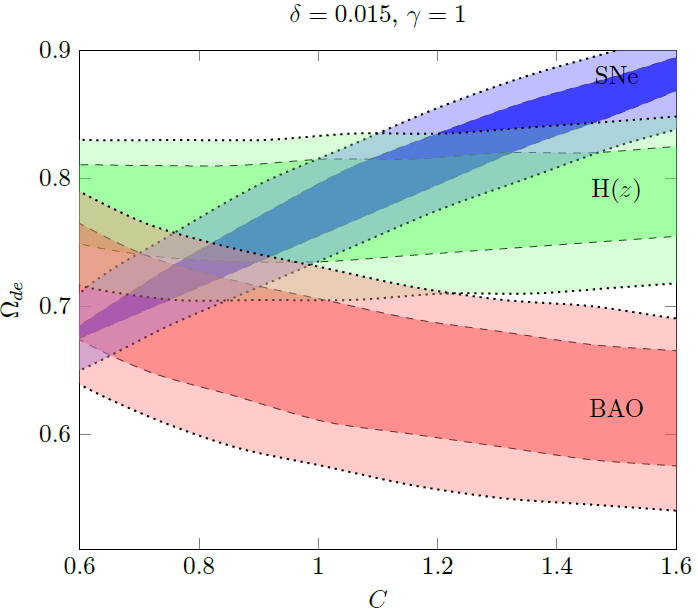}\includegraphics [width=0.35\textwidth]{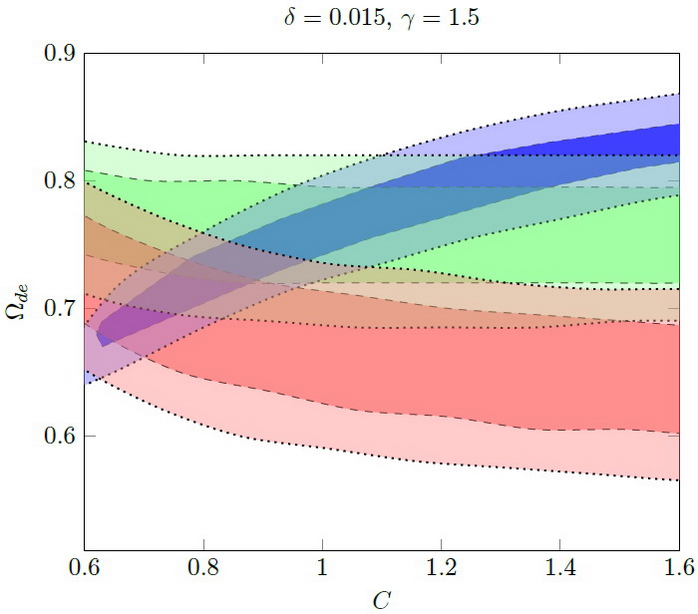}\\
    \caption{$1\sigma$ (dashed lines) and $2\sigma$ (dotted lines) allowed areas on plane $C-\Omega_{de}$ for model of holographic dark energy (\ref{Tsal}) in Friedmann
    cosmology ($\delta=0$) and for Randall-Sundrum brane ($\delta=0.015$). For comparison the results for $\gamma=1$ are given.}
    \label{fig:1}
\end{figure}

We defined $1\sigma$ and $2\sigma$ allowed areas for parameters
$C$ and $\Omega_{de}$ at some fixed values of $\delta$ and
$\gamma$, $\gamma\neq2$ (in this case model simply coincides with
stanadard cosmological $\Lambda$CDM model). For two-parametric
models $68.3\%$ and $95.4\%$ level of likelihood corresponds to
$\chi^2$ for which $\Delta\chi^{2}=\chi ^{2}-\chi_{min}^{2}<2.3$ è
$\Delta\chi^{2}<6.17$ correspondingly. Results of our calculations
for some $\delta$ and $\gamma=1.5$ are given on Fig. \ref{fig:1}.
For $\gamma=1.25$ and $1.75$ similar picture was obtained.

One note that for $\delta\approx 0.02$ we have no joint
intersection between $1\sigma$ and $2\sigma$ allowed regions for
three types of considered data. Therefore we have upper limit on
parameter $\delta \approx 0.02$ for this models and this limit
doesn't depend from value of parameter $\gamma$.

The next question is difference between considered model and
$\Lambda$CDM cosmology. For analysis it is useful to investigate
behavior of so called decceleration parameter
\begin{equation}
q=-\frac{d^2 a}{dt^2}\frac{1}{a H^2}
\end{equation}
and r-parameter
\begin{equation}
r=-\frac{d^3 a}{dt^3}\frac{1}{a H^3}.
\end{equation}
Sahni et al. \cite{Sahni} proposed statefinder pair ($r$,$s$) for
analysis of cosmological evolution where $s$ is
\begin{equation}
s=\frac{r-1}{3(q-1/2)}.
\end{equation}

\begin{figure}[h!]
    \centering
    \includegraphics[scale=0.4]{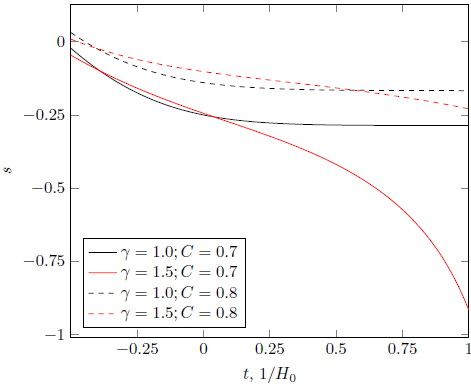}
    \includegraphics [scale=0.4]{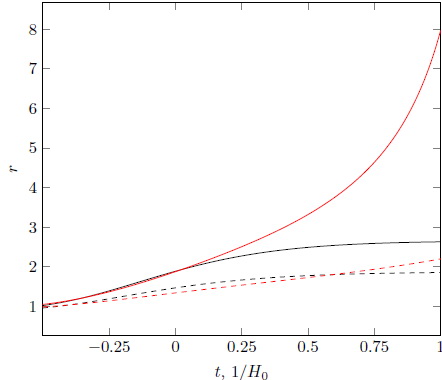}\\
    \includegraphics [scale=0.4]{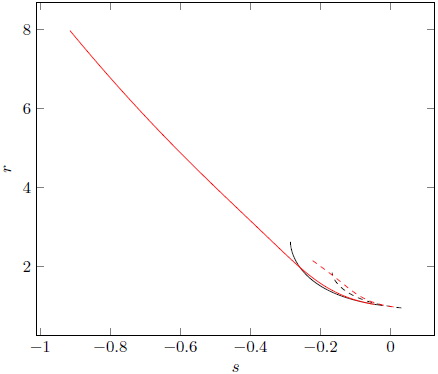}
    \caption{Time evolution of state-finder parameters $r$ and $s$ and corresponding diagram on $(r,s)$-plane for holographic dark
    energy model (\ref{Tsal}) in a case of Friedmann cosmology. Negative values of $t$ corresponds to past, time is given in units of inverse Hubble parameter. Solid and dashed lines
    corresponds to $C=0.7$ and $C=0.8$. The value of $\Omega_{de}$ is assumed 0.72.}
    \label{fig:2}
\end{figure}

\begin{figure}[h!]
    \centering
    \includegraphics[scale=0.4]{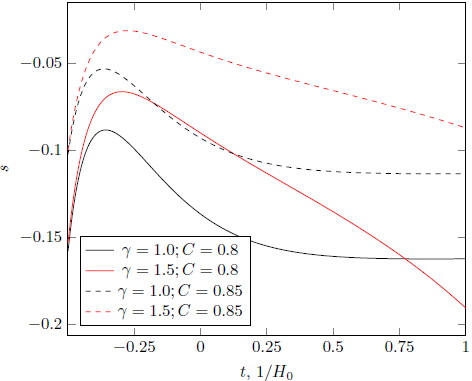}
    \includegraphics [scale=0.4]{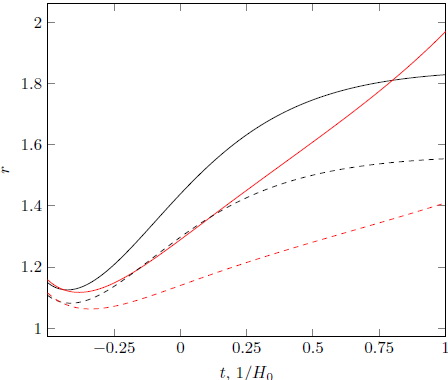}\\
    \includegraphics [scale=0.4]{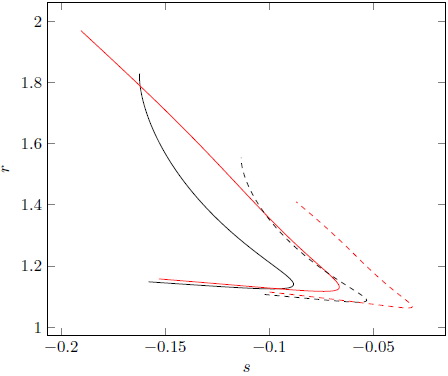}
    \caption{Time evolution of state-finder parameters $r$ and $s$ and corresponding diagram on $(r,s)$-plane for holographic dark energy model (\ref{Tsal}) in a case of
 Randall-Sandrum brane for $\delta=0.015$. Solid and dashed lines
    corresponds to $C=0.8$ and $C=0.85$ on brane. The value of $\Omega_{de}$ is assumed 0.72.}
    \label{fig:22}
\end{figure}

For $\Lambda$CDM model $r=1$ and $s=0$. For considered models the
possible evolution of $s$, $r$ parameters strongly depends from
$C$ and $\gamma$ (see Fig. \ref{fig:2}, \ref{fig:22}). For $C$ and
$\Omega_{de}$ we take values for which the concordance with
observational data is close to $\Lambda$CDM model.

For brane cosmology the value of $s$ is more close to $0$ in
present time in comparison with Friedmann cosmology. For
$\gamma=1$ (standard holographic dark energy model) $s\rightarrow
\mbox{const}$ for brane and Friedmann cosmology.

The deviation of $s$ from $0$ with time grows faster in Friedmann
cosmology. The same one can see for parameter $r$.

One need to note that for some values of $\gamma$ and $C$
state-finder parameters are very close to $\Lambda$CDM values in a
wide range of time $\sim -0.5<t<\sim 1$.

It is interesting to investigate question about future
singularities in considered model. The singularity occurs if
$1<\gamma<2$ both for cosmology on the brane and Friedmann
universe (for values of parameters allowed by observational data
analysis). For given $C$ and $\Omega_{de}$ the time before
singularity increases with increasing of $\delta$ in comparison
with Friedmann cosmology (see table \ref{tab:4}). One also note
that for $1.5\leq\gamma<2$ life of universe ends in so called
``big freeze'' singularity (Hubble parameter diverges for finite
time but scale factor $a=a_{f}\neq \infty$). If $1<\gamma<1.5$ big
freeze singularity occurs only in Friedmann universe while as for
brane big rip take place.

\begin{table}
    \caption{\label{tab:4} Time before singularity $t_{f}$ in model (\ref{Tsal}) for brane ($\delta=0.015$) and Friedmann cosmology ($\delta=0$) for $1\leq\gamma<2$.
    In brackets the type of singularity is given according to classification offered in \cite{NOT}.}
    \begin{center}
        \begin{tabular}{|c|c|}
            \hline Parameters & $t_{f}$, $H_{0}^{-1}$  \\
            \hline $\delta=0$, $\gamma=1$, $C=0.7$, $\Omega_{de}=0.72$ & 2.922 (II)  \\
            \hline $\delta=0$, $\gamma=1.25$, $C=0.7$,
            $\Omega_{de}=0.72$ & 1.488 (III)\\
            \hline $\delta=0$, $\gamma=1.5$, $C=0.7$, $\Omega_{de}=0.72$ & 1.215 (III) \\
            \hline $\delta=0$, $\gamma=1.75$, $C=0.7$, $\Omega_{de}=0.72$ & 0.628 (III) \\
            \hline $\delta=0.015$, $\gamma=1.0$, $C=0.8$, $\Omega_{de}=0.72$ & $\infty$ \\
            \hline $\delta=0.015$, $\gamma=1.25$, $C=0.8$, $\Omega_{de}=0.72$ & 2.652 (II) \\
            \hline $\delta=0.015$, $\gamma=1.5$, $C=0.8$, $\Omega_{de}=0.72$ & 2.287 (III) \\
            \hline $\delta=0.015$, $\gamma=1.75$, $C=0.8$, $\Omega_{de}=0.72$ & 1.594 (III) \\
            \hline
        \end{tabular}
    \end{center}
\end{table}

{\textbf{Remark 1.} We considered model of Tsallis dark energy on
brane but the similar cosmological evolution can be obtained in
frames of generalized holographic dark energy proposed in
\cite{Nojiri}. For universe filled of dark energy only we can put
$$
\rho_{\Lambda}=\frac{\tilde{C}^{2}}{L_{\Lambda}^{2}}=\frac{C^{2}}{L_{e}^{4-2\gamma}}\left(1+\frac{C^{2}}{2\lambda
L_{e}^{4-2\gamma}}\right).
$$
Here $L_{\Lambda}$ is function of $L_{e}$ and parameters $C$ and
$\lambda$. In this case Friedmann equation on the brane take the
form of usual cosmological equation:
\begin{equation}
H^{2}=\frac{\tilde{C}^{2}}{L_{\Lambda}^{2}}\end{equation} }

{Therefore required scale of cut-off is equal
$$
L_{\Lambda}=\alpha
L_{e}^{4-2\gamma}\left(L_{e}^{4-2\gamma}+\beta\right)^{-1/2},\quad
\alpha=\tilde{C}/C,\quad \beta=C^{2}/2\lambda.
$$
}

{One can consider fluid with specific equation of state instead
cosmological model on the brane (see for example
\cite{Timoshkin}).}

{The energy density $\tilde{\rho}$ and pressure $\tilde{p}$ for
equivalent one-fluid Friedmann model mimicking brane cosmology are
$$
\tilde{\rho}=\rho\left(1+\frac{\rho}{2\lambda}\right),
$$
$$
\tilde{p}=p+\frac{\rho}{2\lambda}\left(2p+\rho\right).
$$
One can find that
$$
\frac{d\tilde{p}}{d\tilde{\rho}}=\frac{dp}{d\rho}+\frac{\rho+p}{\rho+\lambda}.
$$

Realistic model requires to account the existence of matter also.
In frames of equivalent Friedmann cosmological model this leads to
the interaction between dark energy and matter. One can obtain the
following expressions for dark energy density and pressure in
Friedmann cosmology in terms of $\rho_{de}$, $\rho_{m}$ and
$p_{de}$ on the brane:
$$
\tilde{\rho}_{de}=\rho_{de}\left(1+\frac{\rho_{de}}{2\lambda}\right)+\frac{\rho_{m}}{2\lambda}(\rho_{m}+2\rho_{de}),
$$
$$
\tilde{p}_{de}=p_{de}\left(1+\frac{\rho_{de}}{\lambda}\right)+\frac{\rho_{de}^{2}}{2\lambda}+\frac{\rho_{m}}{2\lambda}\left(\rho_{m}+2\rho_{de}+2p_{de}\right).
$$
}

{One can find the equation-of-state parameter $w=p_{de}/\rho_{de}$
for considered model on brane and parameter $\tilde{w}$ equivalent
model in Friedmann cosmology. From equation of energy conservation
$$
\dot{\rho}_{de}+3H(\rho_{de}+p_{de})=0
$$
follows that
\begin{equation}
w=-1-\frac{1}{3}(\ln\rho_{de})'.
\end{equation}
Here prime denotes differentiation on $\ln a$. From (\ref{Tsal})
we have
$$
w=-1+\frac{1}{3}\left(4-2\gamma\right)(\ln L_{e})'.
$$
One can rewrite equation for $L_{e}$ as

\begin{equation}
L_{e}=a\int_{a}^{\infty}\frac{da}{Ha^{2}}
\end{equation}
and taking derivative obtain for $w$:

\begin{equation}
w=\frac{1}{3}-\frac{2\gamma}{3}-\frac{1}{3C}(4-2\gamma)\sqrt{\frac{\rho_{de}}{H^{2}}}.
\end{equation}
Or via quantities $\Omega_{de}$ and energy density
$\rho=\rho_{de}+\rho_{m}$:
$$
w=\frac{1}{3}-\frac{2\gamma}{3}-\frac{1}{3C}(4-2\gamma)\sqrt{\frac{\Omega_{de}}{1+\delta
x}},\quad x=\rho/\rho_{0}.
$$
After calculations using equations for $\tilde{\rho}_{de}$ and
$\tilde{p}_{de}$ one obtain for equivalent Friedmann model:
\begin{equation}
\tilde{w}=\frac{w(1+2\delta x\Omega_{de})+\delta
x}{\Omega_{de}+\delta x}.
\end{equation}
}

{If $L_{e}$ decreases and $L_{e}\rightarrow 0$ we have that
x=$\rho/\rho_{0}>>1$ and therefore
$$
w\rightarrow \frac{1}{3}-\frac{2\gamma}{3}
$$
It is interesting to note that parameter $\tilde{w}$ for this case
tends to
$$
\tilde{w}\rightarrow 2w.
$$

In particular for $\gamma \geq 1.25$ $\tilde{w}\rightarrow
\tilde{w}_{f}<-1$ and singularity occurs (see Table \ref{tab:4}).
}

{Of course one can ask, how we can see that considered model is
from braneworld, not just from specific fluid? Of course this
question arises for any cosmological model on the brane. Maybe one
of the arguments in favor of brane cosmology is that this model
consist of relatively simple ingredients in comparison with
complicated form of equation-of-state for cosmological fluid in
Friedmann universe: simple model of holographic dark energy and
simple multidimensional model proposed by Randall and Sundrum.}

{\textbf{Remark 2.} Another interesting question concerns
possibility of complete description of universe history in
considered model including early inflation. For brane we can apply
the approach proposed recently in \cite{Nojiri-2} for Friedmann
cosmology. Namely one put
\begin{equation}
L=\sqrt{L_{e}+\Lambda^{-2}_{UV}},
\end{equation}
where $\Lambda_{UV}$ is some correction due to the ultraviolet
cutoff. For early times the term $\rho/2\lambda>>1$ and
$$
H^{2}\approx \frac{\rho_{2}}{2\lambda}.
$$
Neglecting contribution of matter and radiation we assume that
$\rho=\rho_{de}$ and
\begin{equation}
H\approx \frac{C^{2}}{\sqrt{2\lambda}}\frac{1}{L^{4-2\gamma}}.
\end{equation}
We see for example that for $\gamma=1.5$ this model coincides with
model considered in \cite{Nojiri-2}. Value $C^2/\sqrt{2\lambda}$
plays role of parameter $C$ in \cite{Nojiri-2}. For arbitrary
$0<\gamma<2$ we have similar result i.e. the ultraviolet cutoff
causes exponential expansion at early times.
 }

\section{Another model of  holographic dark energy on the Randall-Sundrum brane}
Let's investigate another model of holographic dark energy model
on brane. One take dark energy density in the following form:
\begin{equation}
\label{eq:25} \rho
_{de}=\frac{C^{2}}{L_{e}^{2}}-{C_{1}^{2}}{H^{2}},
\end{equation}
Here $C_1$ is some constant. Let's consider quantity (energy
density for $t=0$ without $\sim H^{2}$ term):
$$\rho_{0}=\frac{C^{2}}{L_{e}(0)^{2}}+\rho_{m0}.$$
The value of $\rho_{0}$ can be found from equation:
\begin{equation}
\nonumber \left(\rho_{0}-C_{1}^{2} H_{0}^{2}\right )\left (
1+\delta\right)=H_{0}^{2}.
\end{equation}
One use $\Omega_{de}=\rho_{de}/\rho_{0}$ and $C$ as varying
parameters for fixed $C_{1}$ and
$\delta=(\rho_{0}-C_{1}^{2}H^{2})/2\lambda$. The system of
cosmological equations for this model can be written in form (we
again use dimensionless quantities $H/H_0\rightarrow H$, $L_{e}
H\rightarrow L_{e}$, $\rho_{0}/H_{0}^{2}\rightarrow \rho_{0}$):
\begin{equation}
H=\sqrt{rho - C_{1}^{2}\frac{v^{2}}{a^2}}\sqrt{1+\delta
\frac{\rho}{\rho_0-C_{1}^{2}}},
\rho=\frac{C^{2}}{L_{e}^{2}}+\frac{1-\Omega_{de}}{a_{0}^{3}}\rho_{0},
\nonumber
\end{equation}
\begin{equation}
\frac{d{(L_{e}/a)}}{dt}=-\frac{1}{a(t)},
\end{equation}
\begin{equation}
v=\frac{da}{dt}.
\end{equation}

\begin{figure}[h!]
    \centering
    \includegraphics[width=0.35\textwidth]{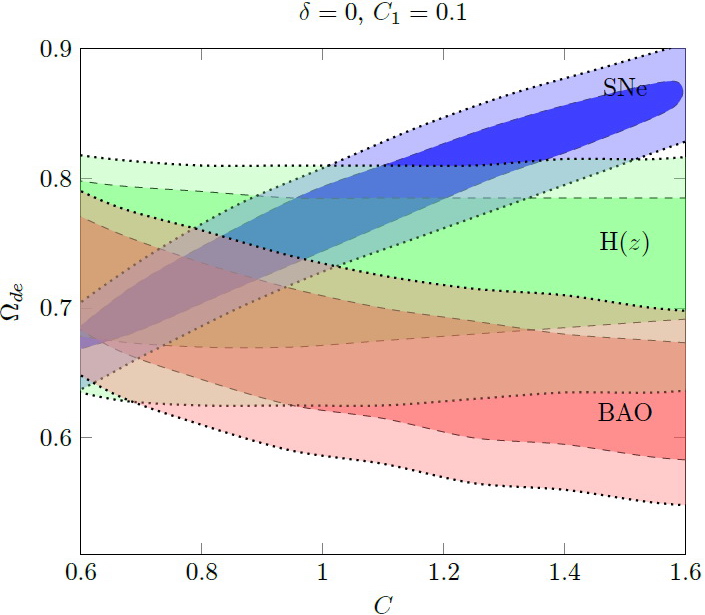}\includegraphics [width=0.35\textwidth]{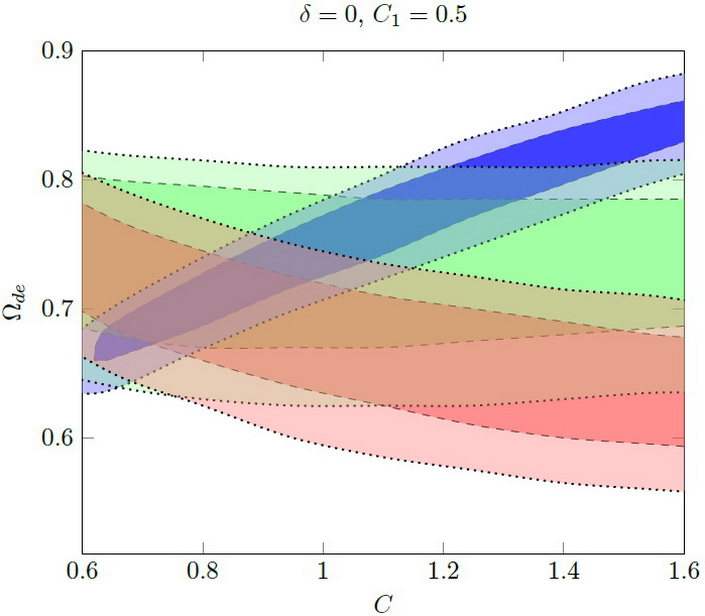}\\
    \includegraphics[width=0.35\textwidth]{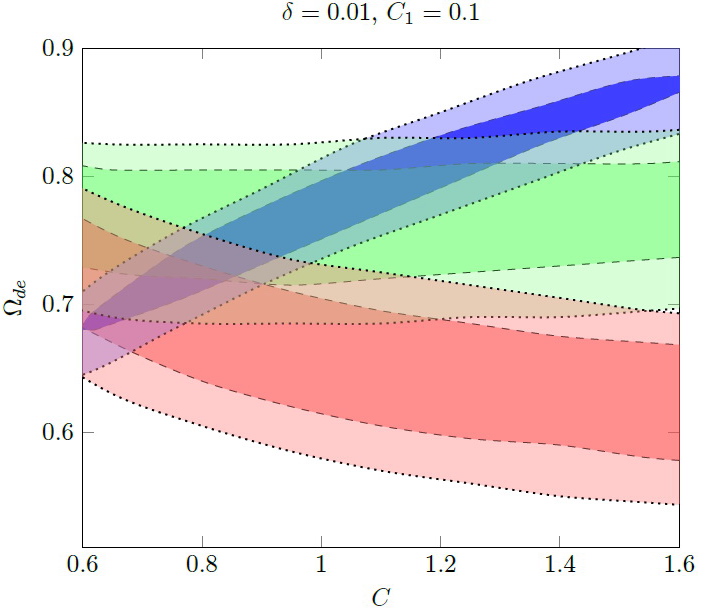}\includegraphics [width=0.35\textwidth]{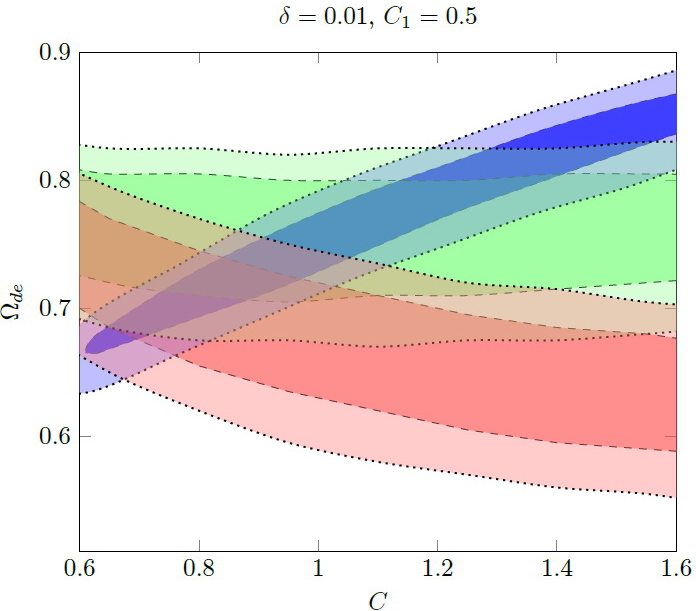}\\
    \includegraphics[width=0.35\textwidth]{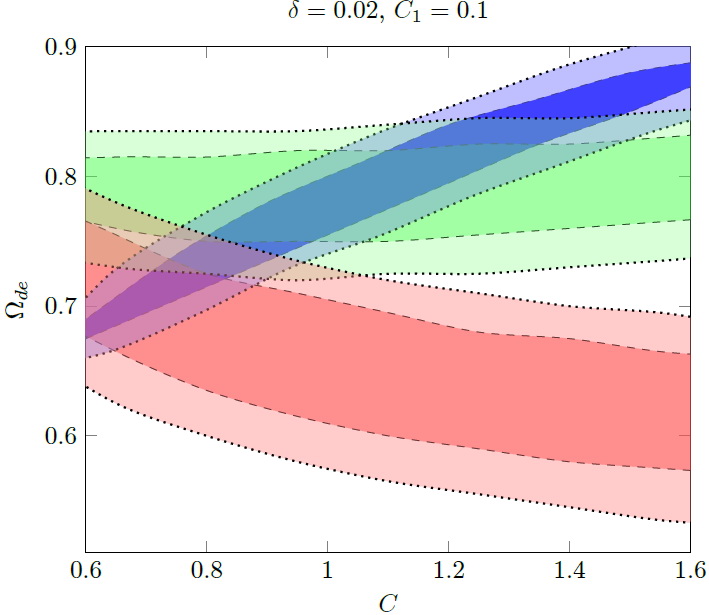}\includegraphics[width=0.35\textwidth]{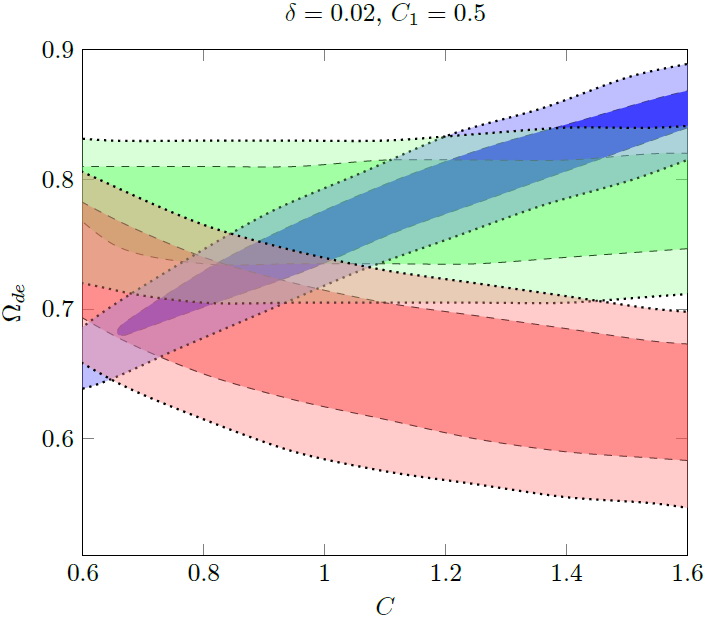}
    \caption{$1\sigma$ (dashed lines) and $2\sigma$ (dotted lines) allowed regions on plane $C-\Omega_{de}$ for model of holographic dark energy (\ref{eq:25}) in Friedmann
    cosmology ($\delta=0$) and for Randall-Sandrum brane ($\delta=0.015$).}
    \label{fig:3}
\end{figure}

Initial conditions are $a(0)=1$ and $v(0)=1$ (latter corresponds
to that the present dimensionless value of Hubble parameter is
simply $1$). For initial value of $L_{e}$ we have via the value of
$\Omega_{de}$:
$$
L_{e}(0)=\left(\frac{C^2}{\Omega_{de}\rho_{0}}\right)^{1/2}.
$$
This model is studied as previous. The allowed $1\sigma$ and
$2\sigma$ regions for parameters $C$ and $\Omega_{de}$ for fixed
values of $\delta$ and $C_{1}$ are given on Fig.\ref{fig:3}.

\begin{figure}[h!]
    \centering
    \includegraphics[width=0.4\textwidth]{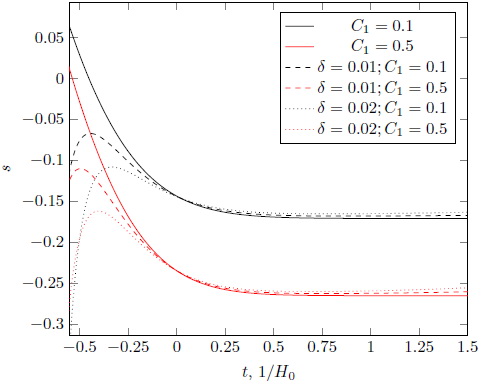}
    \includegraphics [width=0.4\textwidth]{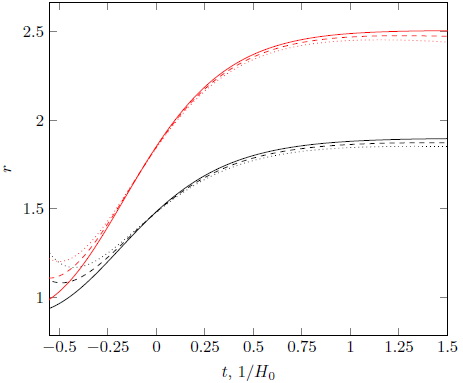}\\
    \includegraphics[width=0.4\textwidth]{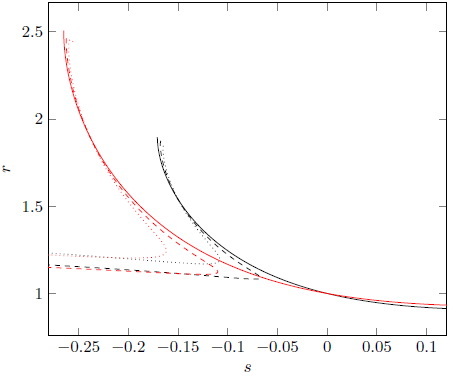}
    \caption{Time evolution of state-finder parameters $r$ and $s$ and corresponding diagram on $(r,s)$-plane for holographic dark energy model (\ref{eq:25})
    in a case of Friedmann cosmology ($\delta=0$) and Randall-Sundrum brane for some $C_1$.}
    \label{fig:4}
\end{figure}

The dependence of state-finder parameters from time are given on
\ref{fig:4}.

One can see as in previous case that at $\delta = 0$ (Friedmann
cosmology) observational data are described better in comparison
with model on the brane. For $\delta \approx 0.02$ and relatively
small values of $C_{1}$ there are no intersections between allowed
areas for various observational data. Therefore we have again some
limit on parameter $\delta$ as in case of Tsallis model. One note
also that in this model there is no final big rip singularity.

\section{Conclusion}

Two classes of holographic dark energy models on Randall-Sandrum
brane are investigated in comparison with Friedmann cosmology and
$\Lambda$CDM model. For first type of models the dark energy
density is assumed to proportional some degree of events horizon
length $\sim L_{e}^{2\gamma-4}$. For second class we presented
dark energy density as sum of two contributions, classical $\sim
L_{e}^{-2}$ and $\sim H^{2}$. Using observational data such as
dependence ``magnitude-redshift'' for SN Ia, dependence of Hubble
parameter from redshift and values of acoustic parameter at some
redshifts one can give allowable areas for parameters of the
models ($\Omega_{de}$ and $C$).

For Tsallis holographic energy model on the brane if $1<\gamma<2$
one cannot avoid singularities in future although time for
singularity increases or singularity became in some sense more
soft (for $\gamma>1.5$ big rip occurs instead singularity of type
III in Friedmann universe).

For second case we there are no singularities in future because
size of events horizon tends to constant value. Such model can be
considered as effective $\Lambda$CDM model in future although
value of ``cosmological contant'' can differs considerably

From analysis of observational data we also obtained the limit on
ratio between current energy density in universe and brane
tension. As follows from our calculations this limit weakly
depends from $\gamma$.

In conclusion one mention about possible perspectives for future
work. One of the important task for example is the reconstruction
of scalar field potential for Tsallis holographic energy on the
brane. Another question is construction of equivalent modified
gravity theory. It is interesting to consider this model in early
universe for loop quantum gravity \cite{Cyclic} also. We are going
to consider these question in the future papers.

The work is supported by project 1.4539.2017/8.9 (MES, Russia).

\newpage

\end{document}